\documentclass[aps,prl,reprint,superscriptaddress,floatfix,longbibliography]{revtex4-2}

\usepackage[
activate={true,compatibility},
final,
tracking=true,
kerning=true,
nopatch=footnote]{microtype}

\usepackage{url}
\usepackage{mathtools}
\usepackage{amsfonts}
\usepackage{nicefrac}
\usepackage{physics}
\usepackage{color}
\usepackage{graphicx}
\usepackage{dcolumn}
\usepackage{booktabs}
\usepackage{tabularx}
\usepackage{multirow}
\usepackage{xcolor}
\usepackage[normalem]{ulem} 
\usepackage{amsthm}
\usepackage{tikz}
\usetikzlibrary{positioning}        
\usepackage[margin=1in]{geometry}
\usepackage{subcaption} 
\usepackage[ruled,vlined]{algorithm2e}
\usepackage{algpseudocode}
\usepackage{hyperref}

\makeatletter
\renewcommand\@makecaption[2]{
  \par
  \vskip\abovecaptionskip
  \begingroup
   \small\rmfamily
    \begingroup
     \samepage
     \flushing
     \let\footnote\@footnotemark@gobble
     \@make@capt@title{#1}{#2}\par
    \endgroup
  \endgroup
  \vskip\belowcaptionskip
}
\makeatother

\newcolumntype{M}{>{\centering}X}
\newcolumntype{Y}{>{\hsize=.125\textwidth\arraybackslash}X}
\newcolumntype{C}{>{\hsize=.25\linewidth\centering\arraybackslash}X}
\newcolumntype{R}{>{\hsize=.125\textwidth\raggedleft\arraybackslash}X}

\begin{document}

\title{Multiscale passive scalar turbulence in a compressed subspace via tensor trains}

\author{Stefano \surname{Pisoni}}
\affiliation{Quantum Research Center, Technology Innovation Institute, Abu Dhabi, UAE}
\affiliation{Institute for Quantum-Inspired and Quantum Optimization, Hamburg University of Technology, Germany}
\author{Egor \surname{Tiunov}}
\affiliation{Quantum Research Center, Technology Innovation Institute, Abu Dhabi, UAE}
\author{Chiara \surname{Calascibetta}}
\email{chiara.calascibetta@inria.fr}
\affiliation{Université Côte d'Azur, Inria, Calisto team, 06902 Sophia Antipolis, France}


\begin{abstract}
Capturing the multiscale statistics of turbulence in compressed form remains a central challenge for reduced-order modeling.  We introduce a hybrid Tensor Train (TT) approach for a highly intermittent passive scalar. The hybrid TT matches Galerkin, wavelet, and standard TT decompositions for the structure functions while improving the representation of intermittent, non-Gaussian fluctuations. These results open a route toward evolving the linear dynamics of passive scalars directly in compressed tensor form, with potential applications to quantum algorithms for fluid transport.
\end{abstract}


\maketitle

\textit{Introduction -- }
Developing compact representations of turbulence without sacrificing its multiscale statistical structure remains a central challenge in computational fluid dynamics~\cite{holmes2012turbulence,moin1998dns}. Long-range correlations, intermittency, and nonlinear interactions across a broad range of scales~\cite{frisch1995turbulence} make it unclear which information must be retained in a compressed representation. Beyond reducing the storage requirements of direct numerical simulations (DNS), answering this question is essential for developing reduced-order solvers operating directly in compressed spaces~\cite{gourianov2022quantum,peddinti,kornev2023chemicalmixer}.
Classical compression strategies retain different features of turbulent flows. Spectral (Galerkin) truncation~\cite{canuto2007spectral} preserves the most energetic Fourier modes, whereas wavelet decompositions~\cite{farge1992wavelet,schneider2010wavelet} retain the largest coefficients of fixed basis functions localized in both space and scale.
Tensor-Train (TT) representations~\cite{oseledets2011tensortrain,Hackbusch_2019} offer a new approach to the problem by encoding high-dimensional signals into a chain of lower dimensional tensors through hierarchical data-driven decompositions.
Originally developed in quantum many-body physics~\cite{White_PRL_1992,Vidal_PRL_2003,schollwock2011density}, TT methods have recently been extended to high-dimensional partial differential equations (PDEs) thanks to the existence of memory efficient TT primitives for differential operators and algebraic manipulations~\cite{gourianov2025tensor,pinkston2025matrix,peddinti,pisoni2025compression,connor2026tensor,compressible_flow_peddinti}. This makes them promising candidates as compressed solvers for turbulent flows.
This perspective naturally identifies passive-scalar transport as the first turbulent benchmark~\cite{celani2000universality}. While retaining the statistical complexity of turbulence (broad inertial range, multiscale correlations, and strong intermittency), its governing equation is linear once the advecting velocity field is prescribed. Passive scalars therefore provide the natural intermediate step between data compression and compressed dynamical evolution, before tackling the non-linear Navier-Stokes equations. Demonstrating that TT representations faithfully preserve the statistical structure of the scalar field is the key prerequisite for such applications ranging from compressed dynamical solvers to quantum algorithms for fluid transport~\cite{jennings2025end_to_end_quantum_for_fluids}. Despite recent applications of TT to fluid-dynamics data~\cite{peddinti,pisoni2025compression}, their ability to capture small-scale intermittency, which controls the extreme fluctuations characteristic of turbulence~\cite{alexakis2018cascades}, remains unclear. 
Most studies have focused on visual reconstruction and low-order statistics~\cite{gourianov2022quantum,quantum-inspired_gpu_acc}, while only limited evidence is available for higher-order observables~\cite{pisoni2025compression}. 
Specifically, in~\cite{pisoni2025compression}, it was shown that the standard TT representation, in which compression is applied uniformly across all scales, may overestimate small-scale fluctuations.
Two related questions therefore remain open. First, can TT representations accurately preserve high-order turbulent statistics, and how do they compare with established Galerkin and wavelet decompositions under identical compression constraints? Second, can alternative tensor-network architectures overcome the limitations of standard TT while retaining the algebraic structure that makes Tensor Trains attractive for compressed dynamical solvers? 
In this Letter, we answer these questions using a highly intermittent passive scalar in the two-dimensional inverse-cascade regime~\cite{calascibetta2025PRF}. Our central technical contribution is a hybrid Tensor Train representation that preserves the coarsest spatial scales exactly while compressing only the finer ones through a standard TT decomposition. We show that this hybrid construction is the key ingredient to improve the representation of deviations from gaussianity of the original fields. 
\begin{figure*}[tbp]
    \centering
    \includegraphics[width=0.95\textwidth]{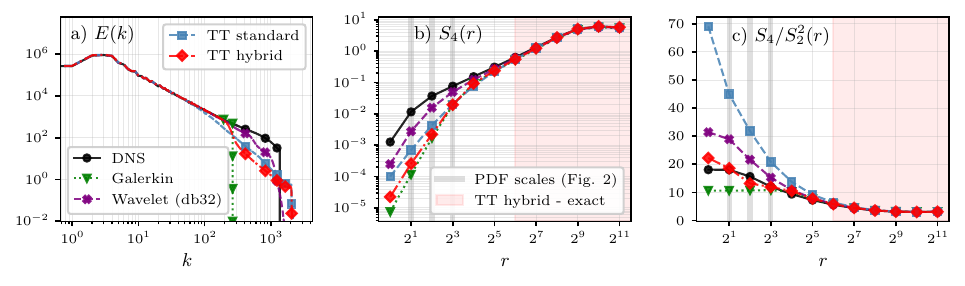}
    \caption{
Comparison of the statistical properties of the passive scalar field reconstructed by Galerkin, wavelet, standard TT, and hybrid TT representations against the DNS at the fixed compression ratio $\rho = 1.31\%$.
\text{(a)} Scalar energy spectrum, $E(k)$. 
\text{(b)} Fourth-order structure function as a function of the separation $r$ (in grid-spacing units), $S_4(r)$.
\text{(c)} Flatness, $F(r) = S_4(r)/S_2^2(r)$.
The shaded region marks the largest scales retained exactly by the hybrid TT ($L_h=6$), while the vertical lines indicate the separations $r=2$, $4$ and $8$ considered in Fig.~\ref{fig:pdf_increments}.
}
    \label{fig:2D_scalar_statistics}
\end{figure*}

\textit{Passive scalar in two-dimensional turbulence --}
 We consider a scalar field $\theta(\mathbf{x},t)$ transported by a two-dimensional incompressible velocity field $\mathbf{u}(\mathbf{x},t)$ evolving in the inverse-cascade regime~\cite{celani2000universality},
\begin{equation}
\partial_t\theta+\mathbf{u}\cdot\nabla\theta
=
\kappa\nabla^2\theta+f_\theta\,,\label{eq:passive_scalar_sca}
\end{equation}
\begin{equation}
\begin{split}
\partial_t\mathbf{u}
+\mathbf{u}\cdot\nabla\mathbf{u}
=
-\nabla p &
+\nu\nabla^2\mathbf{u}
-\beta\mathbf{u}
+\mathbf{f_u}\,,\label{eq:passive_scalar_vel}\\
\nabla\cdot\mathbf{u}&=0\,,
\end{split}
\end{equation}
where $\kappa$ and $\nu$ are the scalar diffusivity and the kinematic viscosity, respectively. The forcing $f_\theta$ injects scalar fluctuations at large scales, whereas $\mathbf{f_u}$ injects kinetic energy at small scales, sustaining the inverse energy cascade. The linear friction $\beta$ prevents energy accumulation at the largest scales.
Although the velocity field displays nearly self-similar statistics~\cite{boffetta2000inverse}, the advected scalar develops the characteristic \emph{ramp-and-cliff} organization~\cite{celani2001fronts}, where extended regions of smooth scalar variation are separated by sharp fronts with intense gradients. These structures generate strong small-scale intermittency: rare and intense scalar gradients dominate high-order statistics while contributing only weakly to the total scalar variance, making passive scalar turbulence a particularly demanding benchmark for any compression strategy.
We consider statistically stationary direct numerical simulations of Eqs.\eqref{eq:passive_scalar_sca}-\eqref{eq:passive_scalar_vel} on a periodic $4096^2$ grid (see Refs.~\cite{calascibetta2025PRF,calascibetta2025turbsca} for numerical implementation details). A representative snapshot of the scalar field is shown in Fig.~\ref{fig:method}a. The statistical complexity of the dataset is summarized by the DNS results (black symbols) in Fig.~\ref{fig:2D_scalar_statistics}, obtained by averaging over an ensemble of 100 statistically independent realizations. The scalar spectrum exhibits the expected inertial-range scaling over more than one decade (Fig.~\ref{fig:2D_scalar_statistics}a). 
To study intermittency, we consider the scalar increments
$
\delta_r\theta=\theta(\mathbf{x}+r\mathbf{e})-\theta(\mathbf{x}),
$
their fourth-order structure function
$
S_4(r)=\left\langle |\delta_r\theta|^4\right\rangle,
$
shown in Fig.~\ref{fig:2D_scalar_statistics}b, and the flatness
$
F(r)=\frac{S_4(r)}{S_2(r)^2},
$
reported in Fig.~\ref{fig:2D_scalar_statistics}c. The flatness is particularly important because it highlights departure from gaussianity. While a Gaussian field has $F=3$, the DNS reaches values close to $F\simeq17$ at the smallest separations, revealing the strong small-scale intermittency. The remaining curves in Fig.~\ref{fig:2D_scalar_statistics} quantify, as discussed later, how accurately each compression method reproduces these observables at a fixed compression ratio. Preserving these highly intermittent small-scale statistics at high compression ratio constitutes the central challenge of this Letter.
\begin{figure*}[tbp]
    \centering
    \includegraphics[width=0.95\textwidth]{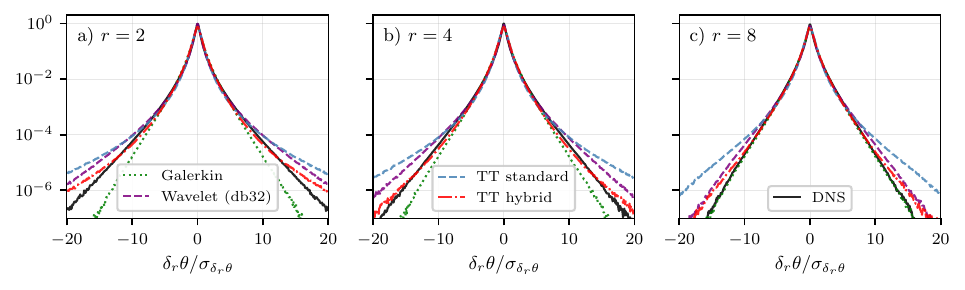}
    \caption{
Probability density functions of the normalized scalar increments at separations a) $r=2$, b) $r=4$ and c) $r=8$.  Galerkin, wavelet, standard TT, and hybrid TT representations, compared with the DNS, at $\rho=1.31\%$.
}
    \label{fig:pdf_increments}
\end{figure*}

\textit{Small-scale intermittent statistics of the compressed fields --} Before describing the technical details of the standard and hybrid TT (next section), we compare four compression strategies: Galerkin~\cite{canuto2007spectral} and wavelet~\cite{farge1992wavelet} truncations, the standard Tensor Train, and the hybrid TT representation at a fixed compression corresponding to approximately $1\%$ of the original DNS parameters. Unlike the standard TT, the hybrid approach preserves the correlations associated with the coarsest spatial scales by construction, while compressing only the finer scales. The red shaded region in Fig.~\ref{fig:2D_scalar_statistics} indicates the scales retained exactly. The representation cost of each compression method is measured by the number of stored parameters $N_{\rm par}$. Throughout this Letter, all methods are compared at identical compression ratio
\begin{equation}\label{eq:rho}
\rho = 100\,\Big(\frac{N_{\rm par}}{N_{\rm DNS}}\Big)\%,
\end{equation}
where the original DNS field on the $4096^2$ grid contains $N_{\rm DNS}=2^{2N}$
degrees of freedom, with $N=12$ binary refinement levels (or scales) in each spatial direction. 
Fixing $\rho$ ensures that differences reflect only the efficiency of the compression strategy rather than the number of retained parameters. In the following we consider a representative compression ratio $\rho=1.31\%$, while the dependence on $\rho$ is discussed later in the Letter. 
Figure~\ref{fig:2D_scalar_statistics} compares the statistical observables reconstructed by the four compression strategies. The scalar spectrum and fourth-order structure function are reproduced with comparable accuracy by all methods (Fig. 1a,b), although the latter is progressively underestimated below $r\approx32$ and with the wavelet decomposition exhibiting the better representation.  
A different picture emerges for the flatness (Fig.~\ref{fig:2D_scalar_statistics}c). The Galerkin truncation suppresses the rapid increase of the flatness, whereas the wavelet representation moderately overestimates it. The standard TT performs even worse, dramatically overestimating the flatness. This shows that blindly applying TT compression across all scales fails to preserve intermittency. In contrast, the hybrid TT remains in close agreement with the DNS.
The differences become even more evident in the probability density functions of the scalar increments shown in Fig.~\ref{fig:pdf_increments}. At the smallest separations the DNS exhibits pronounced non-Gaussian tails associated with the ramp-and-cliff structures~\cite{celani2001fronts}. The hybrid TT better preserves these tails at all three separations ($r=2,4,8$ in grid space units). By contrast, Galerkin truncation underestimates their amplitude, whereas the standard TT and wavelet representations systematically overestimate them, consistent with the behavior observed for the flatness. The origin of these artifacts becomes more clear from the construction of the tensor representations, discussed next.

\textit{Compression methods --}
Galerkin truncation retains Fourier modes below a prescribed cutoff, while wavelet compression preserves the largest coefficients of a Daubechies (db32) decomposition, which captures localized energetic structures~\cite{daubechies1992ten}; the db32 basis was selected as the optimal choice among those tested (e.g., db4 and Haar). Tensor-Train representations instead compress the multiscale correlations of the field. Their construction is summarized in Fig.~\ref{fig:method}, while full details are provided in Appendix~A. 
\begin{figure*}[tbp]
    \centering
    \includegraphics[width=0.85\textwidth]{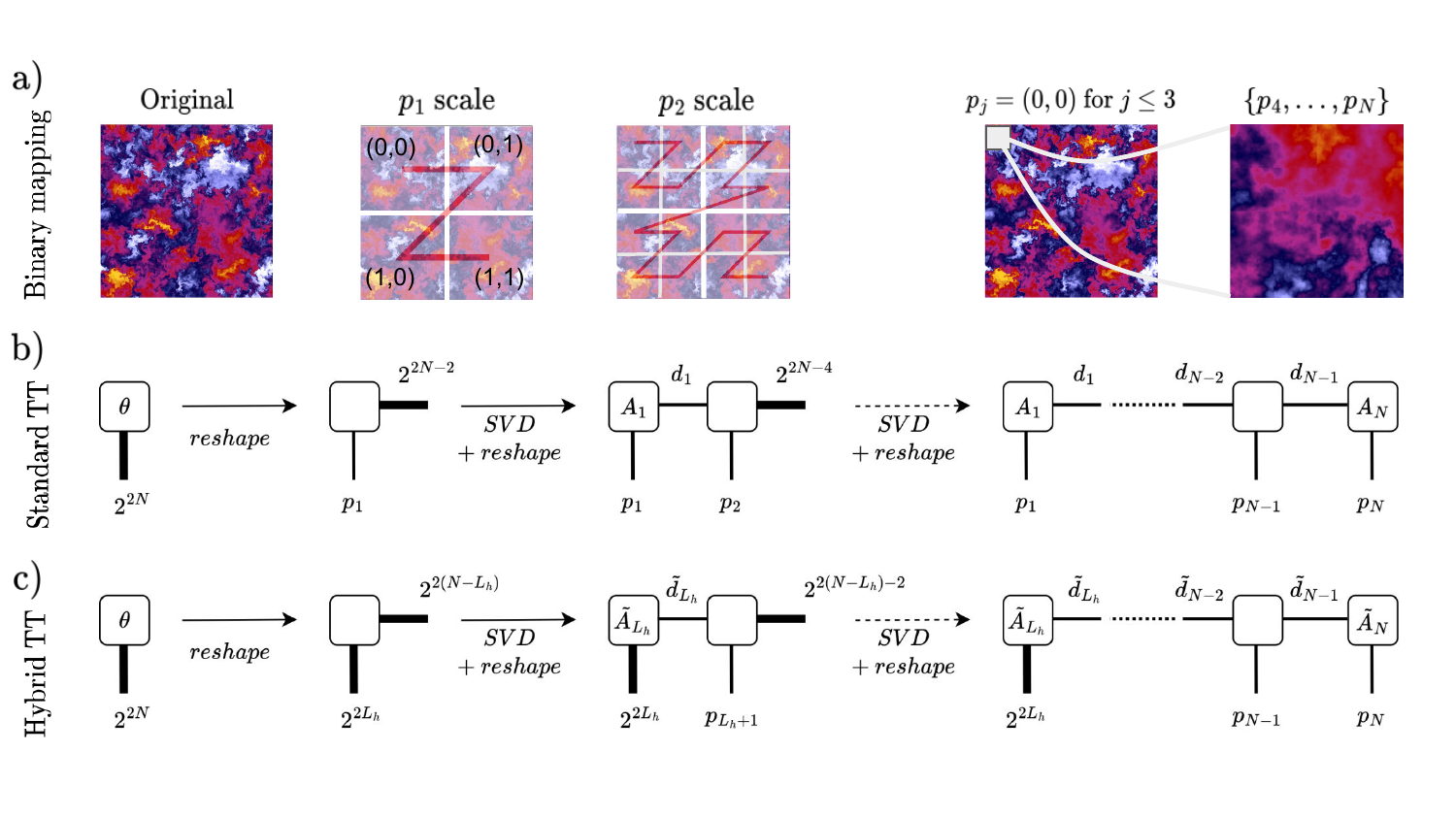}
    \caption{
Tensor Train (TT) representation of a two-dimensional scalar field.
\text{(a)} Binary tensorization through Morton ($Z$-order) indexing. Each physical index $p_j=(x_j,y_j)$ identifies one spatial scale, from the coarsest $p_1$ to the finest $p_N$; successive indices identify progressively smaller regions of the domain.
\text{(b)} Standard TT decomposition. Successive singular value decompositions (SVDs) factorize the tensor into a chain of tensors $A_j$ connected by bond indices $d_j$.
\text{(c)} Hybrid TT decomposition. The first $L_h$ coarsest scales are retained exactly in a dense head tensor, while the remaining $N-L_h$ ones are compressed through a standard TT with bond indices $\tilde d_j$.
}
    \label{fig:method}
\end{figure*}
The $4096^2$ grid is first rewritten through the binary representation of the spatial coordinates. After Morton ($Z$-order) interleaving~\cite{Markeeva2020}, each spatial scale is identified by an index
$p_j=(x_j,y_j)$ (usually called the \textit{physical index} in TT terminology)
(see Fig.~\ref{fig:method}a), with $x_j,y_j \in \{0,1\}  \ \forall j$.
With this prescription, $p_1$ labels the coarsest spatial scale and $p_N$ the finest one. Successive values of $\{p_1,p_2,\ldots\}$ therefore identify progressively smaller regions of the scalar field.
A sequence of singular value decompositions (SVDs)~\cite{oseledets2011tensortrain} then factorizes the resulting tensor into the Tensor Train shown in Fig.~\ref{fig:method}b,
\begin{equation}\label{eq:TT_stand}
\theta \simeq A_{d_1}^{(p_1)}A_{d_1,d_2}^{(p_2)}\cdots A_{d_{N-2}, d_{N-1}}^{(p_{N-1})} A_{d_{N-1}}^{(p_N)},
\end{equation}
where $d_j$'s are defined as the \textit{bond indices} for all $j$'s.
At each refinement level, the SVD orders the multiscale correlations according to their singular values. Compression is then achieved by truncating this hierarchy and imposing
\begin{equation}\label{eq:TT_bond}
d_j\le\chi\,,
\end{equation}
where $\chi$ is the maximal bond dimension. This retains only the dominant correlations while preserving the hierarchical organization of the field.
The resulting storage cost is reduced from $2^{2N}$ to $\mathcal{O}(N\chi^2)$ (see Appendix~A).
The overestimation of intermittency observed in Figs.~\ref{fig:2D_scalar_statistics} and \ref{fig:pdf_increments} originates from this property. 
In fact, the TT compression strategy globally smoothens the scalar field, but discontinuities aligned with the grid appear at each scale refinement due to the hierarchical nature of the encoding. These discontinuities become significant compared to the smoothened field, leading to artificial extreme events already observed in~\cite{pisoni2025compression} at small bond dimensions.
We introduced the {hybrid TT} (Fig.~\ref{fig:method}c), in which the first $L_h$ coarsest scales ($h$ for head) are retained exactly in a dense head tensor, while only the remaining $N-L_h$ finer scales are compressed through a standard TT with maximal bond dimension $\tilde{\chi}$. Throughout this Letter, we fixed $L_h=6$, corresponding to the red shaded region in Fig.~\ref{fig:2D_scalar_statistics}; the dependence on $L_h$ is discussed in Appendix~B.
The dense head partitions the domain into known $2^{L_h}\times 2^{L_h}$ blocks, localizing the compression-induced discontinuities at their interfaces. These artifacts can therefore be mitigated through a local linear interpolation between neighbouring grid points.
In all the results presented above, hybrid TT refers to this interpolation-corrected representation.

\begin{figure}[tbp]
    \centering
    \includegraphics[width=0.85\columnwidth]{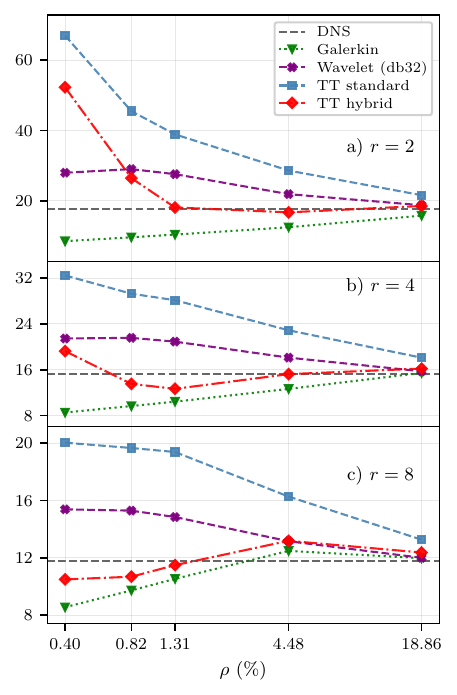}
    \caption{
Flatness $F(r)=S_4/S_2^2$ at separations a) $r=2$, b) $4$ and c) $8$ as a function of the compression ratio $\rho$. The dashed gray lines denote the DNS reference. 
}\label{fig:varying_bond_dimension}
\end{figure}

\textit{Compression-ratio dependence --}
We finally assess the robustness of the previous observations by varying the compression ratio. The nominal TT bond dimensions considered in the following are $\chi=\{50,75,100,200,500\}$; the corresponding Galerkin, wavelet, standard and hybrid TT representations are calibrated to the same value of the compression ratio (Eq.\eqref{eq:rho}), namely $\rho \in \{0.40\%, 0.82\%, 1.31\%, 4.48\%, 18.86\%\}$.
Figure~\ref{fig:varying_bond_dimension} reports the flatness at separations $r=2$, $4$ and $8$ as a function of $\rho$.
At strong compressions, hybrid TT shows the best overall agreement with the DNS, whereas Galerkin underestimates intermittency and wavelet/standard TT overestimate it.
As the compression ratio increases, all methods approach the DNS, although from opposite directions: Galerkin converges from below, reflecting the loss of intense scalar gradients, whereas the hybrid TT converges from above, Fig.~\ref{fig:varying_bond_dimension}. The main advantage of the hybrid TT therefore lies in the strongly compressed regime, where representing intermittent statistics is most demanding. These results demonstrate that preserving the coarsest multiscale correlations before compressing the finer scales improves the representation of turbulent intermittency.

\textit{Discussion --}
We have systematically compared Fourier, wavelet, Tensor Train representations and proposed a hybrid TT for the compression of a highly intermittent passive scalar under identical compression ratios. While both standard and hybrid TT accurately reproduce the large-scale structures of the flow, only the hybrid TT faithfully preserves the small-scale flatness, suppressing the compression-induced discontinuities that affect the standard TT representation.
More importantly, unlike Galerkin and wavelet decompositions, tensor-network representations adapt to the data correlations while retaining an algebraic structure that naturally supports compressed dynamical solvers~\cite{peddinti}. Furthermore, the hybrid TT representation naturally suggests hybrid numerical solvers, where the dense head tensor is evolved using sparse linear-algebra techniques while the remaining compressed hierarchy is propagated through standard Tensor-Train algorithms. Rather than reconstructing the full scalar field after each time step~Eq.\eqref{eq:passive_scalar_sca}, one could evolve its Tensor Train representation by applying the discrete advection-diffusion operator in compressed form. This formulation naturally connects to Matrix Product Operators and, ultimately, to quantum algorithms based on block-encoded evolution operators~\cite{pisoni2025compression,SieglPRR_2026}.
The Kraichnan model of passive scalar turbulence provides the natural next step in this direction~\cite{kraichnan1994anomalous}. While the present work considers passive scalar advected by a velocity field obtained from direct numerical simulations of the Navier-Stokes equations, Eqs.~\eqref{eq:passive_scalar_vel}~\cite{calascibetta2025PRF,calascibetta2025turbsca}, the Kraichnan model replaces the advecting field with a prescribed Gaussian, $\delta$-correlated velocity field. This retains the linear transport equation~Eq.\eqref{eq:passive_scalar_sca} while introducing statistically independent advection operators, making the Kraichnan model an ideal bridge toward compressed dynamical solvers based on tensor-network representations of the passive scalar system.
Moreover, the Kraichnan model admits analytical predictions for anomalous scaling in well-controlled asymptotic regimes (e.g., in the limits of very smooth and very rough advecting velocity fields), making it particularly well suited to investigate how the tensor-network complexity required to achieve a prescribed accuracy depends on intermittency. Such research direction bridges multiscale turbulence, tensor-network methods, and quantum algorithms. 

Overall, this work establishes hybrid tensor trains as a viable framework for compressed passive scalar turbulence representations, laying the foundation for compressed fluid transport solvers using both classical settings and quantum algorithms.

\textit{Acknowledgements --}
The authors thank Luca Biferale, Michele Buzzicotti, Raghavendra Peddinti, Leandro Aolita, and Maurizio Carbone for useful discussions. 
\bibliographystyle{unsrt}
\bibliography{biblio}

\begin{center}
\textbf{END MATTER}
\end{center}

\textit{Appendix A: Tensorization and Tensor Train representation --} This appendix provides additional details on the tensorization procedure underlying the Tensor Train representation introduced in the main text.
The passive scalar $\theta(x,y)$ is defined on a Cartesian grid containing $2^N\times2^N$ points. Each grid point is identified by two integer coordinates, $x,y \in\{0,\dots\,,2^{N}-1\}$. Every coordinate can be written uniquely using $N$ binary digits,
\[
    x=\sum_{j=1}^{N}{2^{N-j}x_j}, 
\qquad
 y= \sum_{j=1}^{N}{2^{N-j}y_j}\,,
\]
where $x_i,y_i\in\{0,1\}$. The binary digits naturally define a hierarchy of spatial scales. The first digit determines whether the point lies in the left or right half of the domain, the second digit further subdivides that half, and each additional digit doubles the spatial resolution. Consequently, the most significant bits describe the coarsest spatial scales, while the least significant bits describe the finest ones.
To encode the two spatial coordinates into a hierarchical, scale-ordered representation, the binary digits are interleaved according to the Morton (Z-order) indexing~\cite{Markeeva2020},
\[
p=x_1y_1x_2y_2\cdots x_Ny_N\,.
\]
Each pair $(x_j, y_j)$ contains the information associated with the $j-$th refinement level and therefore defines a local (or \textit{physical}) index, 
\[
p_j=(x_j,y_j),
\qquad j=1,\ldots,N\,.
\]
Each physical index can assume four possible values, 
\[
    (0,0),\ (0,1),\ (1,0),\ (1,1)\,,
\]
corresponding to the four quadrants generated at that refinement level (Fig.~3a). The first index $p_1$ identifies one of the four largest quadrants of the domain, the second index specifies one of the four sub-quadrants inside it, and so on until the finest spatial resolution is reached. 
Consequently, the original order-2 tensor $\theta(x,y)$ is reinterpreted as a tensor of order $N$,
\[
\theta(x,y) \equiv \theta(p_1,p_2,\ldots,p_N)\,.
\]
Then, applying a sequence of singular value decompositions (SVDs)~\cite{oseledets2011tensortrain} to this higher-order tensor, one finds Eq.~\eqref{eq:TT_stand}.
The procedure starts by separating the first tensor index from all the remaining ones,
\[
(p_1)\,|\,(p_2,p_3,\ldots,p_N),
\]
which reshapes the tensor $\theta(p_1,p_2,\ldots,p_N)$ into a matrix.
The first SVD then identifies the dominant correlations between the index $p_1$ and the remaining tensor indices and decomposes the matrix into the first TT tensor, $A_{d_1}^{(p_1)}$, and a remainder tensor
$
R_{d_1}^{(p_2,\ldots,p_N)},
$
connected through the bond index $d_1$. The remainder tensor is then reshaped into a proper matrix to which a second SVD is applied, producing the second TT tensor $A_{d_1,d_2}^{(p_2)}$. Iterating this procedure recursively yields the complete Tensor Train representation~\eqref{eq:TT_stand} (Fig.~3b).
At each SVD step, the reshaped matrix is decomposed into a set of orthogonal correlation modes ranked by their singular values. The magnitude of the singular values quantifies the strength of the correlations between the successive refinement levels. Retaining only the largest singular values yields a low-rank approximation, with rank equal to the bond dimension $d_j$ in Eq.~\eqref{eq:TT_bond}. Overall, each internal TT tensor $A_j$ therefore contains one physical index $p_j$, associated with the $j$-th refinement level, and two bond indices, $d_{j-1}$ and $d_j$. 
Its dimensions are therefore
\[
d_{j-1}\times4\times d_j.
\]
The factor four originates from the four possible values of the physical index $p_j=(x_j,y_j)$.  If the bond dimensions are all of the same order, $d_j\sim\chi$, the total number of stored parameters scales as
\[
N_{\rm par}^{\rm TT}
\simeq
4N\chi^2,
\]
up to negligible boundary contributions from the first and last tensors. This expression corresponds to the storage cost reported in the main text.

For the hybrid TT, the first $L_h$ scales are grouped into a single dense head tensor rather than being factorized through successive SVDs. Since this tensor is connected to the remaining TT chain through a bond of dimension $\tilde\chi$, it contains
$2^{2L_h}\tilde\chi$ parameters.
The remaining $N-L_h$ scales are compressed through a standard TT decomposition, contributing
$\left(4(N-L_h)\tilde\chi^2\right)$
parameters. The total parameter cost therefore scales as
\[
N_{\rm par}^{\rm hyb}
\simeq
2^{2L_h}\tilde\chi
+
4(N-L_h)\tilde\chi^2.
\]

In all the results shown in this Letter, $\tilde\chi$ has been chosen so that the hybrid TT had the same compression ratio $\rho$ as the corresponding standard TT.


\textit{Appendix B: Dependence on the parameter $L_h$.}
\begin{figure}[h!]
    \centering
    \includegraphics[width=0.9\columnwidth]{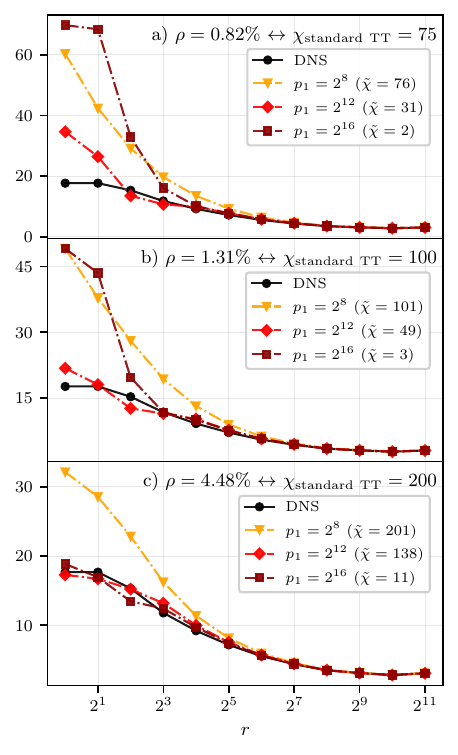}
    \caption{
    Dependence on the hybrid head dimension. Flatness $F(r)$ versus separation $r$ for hybrid TT with head sizes $2^{2L_h}$, $L_h=4,6,8$ (orange triangles, red rhombus, dark red squares), compared with the DNS (black circles). Each panel fixes the compression ratio on the basis of the standard-TT bond dimension, $\chi=75,100,200$ (a-c); the legend reports the corresponding remaining bond dimension $\tilde\chi$ of the compressed chain.}
\label{fig:varying_bhybrid_dimension}
\end{figure}
Figure~\ref{fig:varying_bhybrid_dimension} reports the flatness isolating the role of the hybrid head dimension, $L_h$.
For three fixed compression ratio, $\rho = \{0.82\%, 1.31\%, 4.48\%\}$,  (corresponding to $\chi=75,100,200$), we vary the number of explicitly retained scales through the head size $2^{2L_h}$, $L_h=4,6,8$. 
Increasing $L_h$ moves more of the coarse-scale information into the uncompressed head. However, at a fixed parameter budget this comes at the expense of the maximal remaining bond dimension $\tilde\chi$, which must shrink to keep the total cost constant, degrading the resolution of the finest scales. The hybrid head dimension is thus a genuine control knob, trading the number of explicitly prescribed scales against the fidelity of the compressed remainder. Indeed, the intermediate choice $L_h=6$ used in the main figures is found to be the optimal choice from Fig.~\ref{fig:varying_bhybrid_dimension}. In all the subplots a local interpolation was performed at the hybrid head scale.
\end{document}